\documentclass{emulateapj}
\begin{document}

\title{ Damped Ly$\alpha$ Gas Metallicities  at
Redshifts $z=0.9-2.0$ from SDSS Spectra\altaffilmark{1}}

\author{Daniel B. Nestor\altaffilmark{2}, Sandhya
M. Rao\altaffilmark{2}, David A. Turnshek\altaffilmark{2}, and
Daniel Vanden Berk\altaffilmark{2}}

\affil{Department of Physics \& Astronomy, University of Pittsburgh,
Pittsburgh, PA 15260}

\altaffiltext{1}{Based on data obtained in the Sloan Digital Sky 
Survey and by the Hubble Space Telescope,
operated by STScI-AURA, for NASA.}

\altaffiltext{2}{email: dbn@phyast.pitt.edu,
rao@everest.phyast.pitt.edu,  turnshek@pitt.edu, danvb@phyast.pitt.edu}

\begin{abstract}

Using Sloan Digital Sky Survey (SDSS) early data release spectra,
we have identified 370 \ion{Mg}{2} absorption systems with \ion{Mg}{2}
$\lambda$2796 rest equivalent widths $\ge 1$ \AA\ and redshifts
$z=0.9-2.2$. From our previous and ongoing HST UV spectroscopic
studies, we estimate that the mean neutral hydrogen column
density of a system selected in this manner is $N_{HI} = 3.6\pm
1.3 \times10^{20}$ atoms cm$^{-2}$, which corresponds to the damped
Ly$\alpha$ (DLA) regime. We have formed high signal-to-noise ratio
composite spectra using 223 of these systems with $z=0.9-2.0$
in order to study the strength of the \ion{Zn}{2} and \ion{Cr}{2}
absorption lines corresponding to this mean neutral hydrogen column
density.  After making a correction for missed DLAs, overall we find
that [Zn/H] = $-1.13 \pm 0.19$. We find [Cr/Zn] = $-0.45 \pm 0.13$, which
indicates that $\approx$ 65\% of the Cr is depleted on to grains, but
this does  not correct for the missed DLAs.
We have also derived Zn and Cr abundances in two kinematic regimes,
and within each regime we consider two redshift intervals.  We find
trends which indicate that metallicities are higher in the composites
where the absorption has larger velocity spreads as measured by
\ion{Mg}{2} $\lambda$2796 rest equivalent width.  Larger velocity
spreads may correspond to deeper gravitational potential wells
which represent more massive and chemically evolved structures,
and/or regions associated with winds from starbursting galaxies,
also leading to kinematically broad structures of chemically enriched
gas. Within the large velocity spread regime, we find that at lower
redshifts the Zn metallicity is larger and more Cr is depleted on
to grains.  \end{abstract}


\section{Introduction}


Over a decade ago Pettini and collaborators presented the first
in a series of studies aimed at deriving metal abundances for the
neutral-gas-phase component of the Universe at high redshift (see
Pettini et al. 1999 and references therein). In their initial studies
they relied on medium-resolution spectroscopy of the \ion{Zn}{2}
$\lambda\lambda$2026,2062, and \ion{Cr}{2} 
$\lambda\lambda$2056,2062,2066 absorption lines in QSO 
damped Ly$\alpha$ systems (DLAs).
Measurements of metals in DLAs with $N_{HI} \ge 2\times10^{20}$
atoms cm$^{-2}$ are appropriate for studies of the cosmic
neutral-gas-phase metallicity because these systems trace the bulk
of the neutral hydrogen gas mass of the Universe out to at least
redshift $z\approx3.5$ (see Storrie-Lombardi \& Wolfe 2000, Rao \&
Turnshek 2000, hereafter RT2000, and references therein).  It was
realized that measurements of the \ion{Zn}{2} and \ion{Cr}{2} lines
using medium-resolution spectra would likely be sufficient
to consider the problem because the lines would be unsaturated even in
DLAs.

However, selection effects and small number statistics continue to
be important concerns in these studies.  For example, the presence of
dust in a subset of absorbers would dim the corresponding background
QSOs, causing a bias against finding absorbers with high dust
content.  Moreover, while a column density-weighted determination
of the cosmic metallicity is the most relevant measurement for
tracking the overall chemical evolution of the neutral gas, this
leads to a result mostly dominated by a few of the highest $N_{HI}$
systems. Since individual absorbers have a range of metallicities,
the uncertainty introduced by the small number of very high $N_{HI}$
absorbers studied to date is a problem.

Instead of measuring individual QSO absorption-line systems,
we take the approach of using composite spectra here, which
leads to preliminary measurements of the DLA gas metallicity of
the Universe at moderate redshift.  We derive the composites from
Sloan Digital Sky Survey (SDSS; York et al. 2000) early data release
(EDR; Stoughton et al. 2002) QSO spectra (Schneider et al. 2002),
which individually show evidence for high-$N_{HI}$ absorption based
on their \ion{Mg}{2} absorption-line properties.  By using these
composite spectra, the worries related to small number statistics
can be mitigated.  We have used 223 absorption-line systems in
the redshift interval $0.9 \le z \le 2.0$ to make composites
in this initial study, representing a total integrated neutral
hydrogen column density of $N^{tot}_{HI} \approx 8\times10^{22}$
atoms cm$^{-2}$.  This exceeds corresponding numbers for previously
studied individual systems: 42 systems over the redshift interval
$0.4<z<3.5$ with $N^{tot}_{HI} \approx 4\times10^{22}$ atoms
cm$^{-2}$ (Pettini et al. 1999; Prochaska \& Wolfe 1999; Molaro et
al. 2000; Ge, Bechtold, \& Kulkarni 2001). 

\section{Analysis}

\subsection{Sample Definition}

Using Hubble Space Telescope (HST) UV spectroscopy, RT2000 showed
that strong intervening \ion{Mg}{2}$-$\ion{Fe}{2} absorption-line
systems identified in QSO spectra characteristically have very
high neutral hydrogen column density gas (generally $10^{19} <$
$N_{HI}$ $< 10^{22}$ atoms cm$^{-2}$).  Many are DLAs with $N_{HI}
\ge 2\times10^{20}$ atoms cm$^{-2}$.  RT2000 showed that $\approx
95$\% of these DLAs can be identified by selecting systems which
have a \ion{Mg}{2} $\lambda$2796 rest equivalent width (REW)
$W^{\lambda2796}_0 > 0.5$ \AA\ {\it and} \ion{Fe}{2} $\lambda$2600
REW $W^{\lambda2600}_0 > 0.5$ \AA.  We have used these results to
establish criteria for selecting high-$N_{HI}$ systems in the past.
Here we use a slightly revised criterion ($W^{\lambda2796}_0 \ge
1.0$ \AA) to obtain a sample of 370 high-$N_{HI}$ systems. 
It is straightforward to select absorption-line systems above this
threshhold using SDSS spectra.\footnote{We note that our composite
spectra indicate that $W^{\lambda2796}_0$ is $\approx 1.9$ times
larger than $W^{\lambda2600}_0$.  Thus, we are generally consistent
with the criterion that $W^{\lambda2600}_0 > 0.5$ \AA.} Recognizing
weaker systems is also possible, but with greater 
incompleteness due to signal-to-noise considerations. In Nestor
et al. (2003, in preparation) we will present the findings of a
survey for \ion{Mg}{2} $\lambda\lambda$2796,2802 absorption-line
systems in SDSS QSO spectra. The results will establish the
incidence of \ion{Mg}{2} systems as a function of REW and also
include measurements of \ion{Fe}{2} $\lambda$2600 and \ion{Mg}{1}
$\lambda$2852, when available.  The details of selecting
systems (e.g., continuum fitting, equivalent width measurements,
completeness) will be discussed there.  It should
be emphasized that our selection method basically amounts to a
kinematic, as opposed to a metallicity-based, criterion. This is
because the \ion{Mg}{2} $\lambda$2796 absorption lines are on the
saturated part of the curve of growth, so their REWs are essentially
measures of velocity spread rather than column density.

\subsection{Forming and Measuring the Composite Spectra}

The 370 absorbers in our sample were found in 3683 SDSS EDR
spectra. To form composites
we continuum-normalized each spectrum (and error array) and
wavelength-shifted them into the absorber rest frame. To accomplish
this, a spectrum was first rebinned on a finer subpixel grid,
with subpixels having one-fifth the size of an original pixel; this
minimized data smoothing due to rebinning.  
We then used visual inspection to eliminate 70 spectra that
suffered from abnormalities or poor signal-to-noise in
the \ion{Zn}{2}$-$\ion{Cr}{2} region of interest. In principle,
the remaining 300 spectra would all be valuable for forming
composites. However, one of the goals of our analysis is to make
an accurate measurement of the \ion{Zn}{2} $\lambda2026$ absorption
line, and in SDSS medium-resolution spectra this line is potentially
blended with the low-oscillator-strength \ion{Mg}{1} $\lambda2026$
absorption line. Therefore, in this initial study, we
decided to estimate the contribution of \ion{Mg}{1} $\lambda2026$
to the blend by only using spectra that also had data on the
higher-oscillator-strength (but generally unsaturated) \ion{Mg}{1}
$\lambda2852$ absorption line. This resulted in the elimination of
77 more systems for which \ion{Mg}{2} $\lambda$2852 was unavailable.
We then experimented with different methods for forming composites
and found that inverse variance-weighting produced the highest
signal-to-noise ratio composites, so we adopted results produced
by this method since there is no known correlation between $N_{HI}$
and spectrum signal-to-noise ratio.  Other methods (e.g., straight
averaging) would have produced somewhat lower signal-to-noise
ratio composites.  We note that, if QSO dimming due to the presence of
dust and, therefore, high metal content, is significant, we would
be underestimating the mean metallicity, as our weighting method
favors brighter QSOs.

The composite spectrum in the \ion{Zn}{2}$-$\ion{Cr}{2} region
derived from the entire sample of 223 absorbers is shown in the left
panel of Figure 1.  The right panel of Figure 1 shows other composites
formed by roughly dividing the sample into quarters while keeping
the variance approximately the same. These subsamples correspond
to \ion{Mg}{2} absorbers with  low REW ($1.0 \le W_0^{\lambda2796}
< 1.3$ \AA) and high REW ($W_0^{\lambda2796} \ge 1.3$ \AA), and
at lower redshift ($0.9 \le z < 1.35$) and higher redshift ($1.35
\le z \le 2.0$) within the total sample.  This subdivision leads
to some important results (\S3 and \S4). 
Measurements of the absorption lines in the \ion{Zn}{2}$-$\ion{Cr}{2}
region in the five  composites shown in Figure 1  are reported in
Table 1, along with the characteristic properties that define the
total sample and four subsamples.

\begin{figure*}
\plottwo{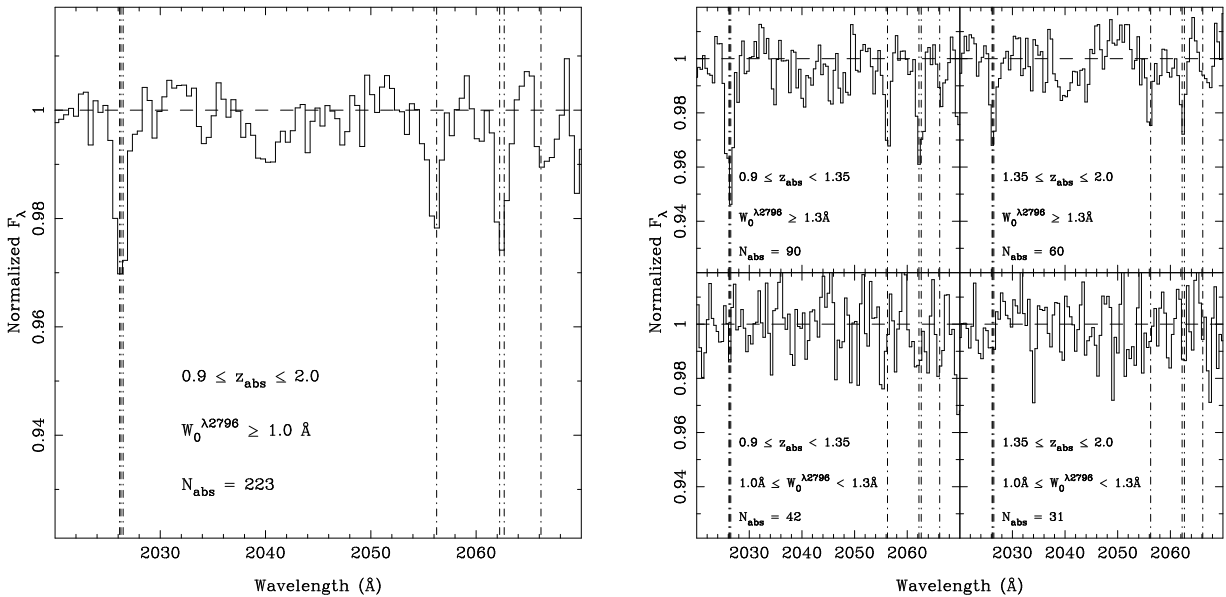}{fig02.eps}
\caption{{\bf Left:} Composite spectrum in the
rest frame \ion{Zn}{2}-\ion{Cr}{2} region derived from 223 SDSS
QSO spectra. The dash-dotted lines indicate the rest frame
locations of the \ion{Zn}{2}, \ion{Cr}{2}, and \ion{Mg}{1} lines
(Table 1). {\bf Right:} Composite spectra for four subsamples of
the total sample divided according to redshift and \ion{Mg}{2}
$\lambda$2796 REW.  Notice the presence of stronger lines
(indicating more chemically evolved structures) in the larger REW
(i.e. larger velocity spread) subsamples, the stronger \ion{Zn}{2}
(indicating increased metallicity) at lower redshift, and the
stronger \ion{Cr}{2} relative to \ion{Zn}{2} (indicating less dust)
at higher redshift.}
\end{figure*}

\begin{deluxetable*}{lrrrrrr}
\tablewidth{0pc}
\tablecaption{Results from Composite Spectra Measurements}
\tablehead{
\colhead{} &
\colhead{Full} &
\colhead{Low-$z$} &
\colhead{High-$z$} &
\colhead{Low-$z$} &
\colhead{High-$z$} &
\colhead{Cosmic}\\[.2ex]
\colhead{Property} &
\colhead{Sample} &
\colhead{High-REW} &
\colhead{High-REW} &
\colhead{Low-REW} &
\colhead{Low-REW} &
\colhead{DLA Gas} \\[.2ex]
\colhead{} &
\colhead{} &
\colhead{Subsample} &
\colhead{Subsample} &
\colhead{Subsample} &
\colhead{Subsample} &
\colhead{} \\[.2ex]
}
\startdata
Number of Absorbers   & 223        & 90         & 60         & 42         & 31          & \nodata \\
\ion{Mg}{2}$\lambda$2796 REW (\AA)  & $\ge1.0$   & $\ge$1.3     & $\ge$1.3     & 1.0$-$1.3  & 1.0$-$1.3   & $>0$ \\
Redshift (z) Interval  & 0.9$-$2.0  & 0.9$-$1.35 & 1.35$-$2.0 & 0.9$-$1.35 & 1.35$-$2.0  & 0.9$-$2.0 \\
\ion{Zn}{2}$\lambda$2026bl$^1$ REW (m\AA)  & 33.6 (5.0) & 70.2 (9.3) & 29.4 (7.7) & 0.1  (10.6)& 6.5  (11.3) & \nodata \\
\ion{Mg}{1}$\lambda$2026bl$^1$ REW (m\AA)  & 10.2 (0.2) & 13.7 (0.3) & 10.9 (0.3) & 5.6  (0.4) & 4.0  (0.5)  & \nodata \\
\ion{Cr}{2}$\lambda$2056 REW (m\AA)  & 31.1 (5.1) & 41.7 (7.7) & 32.2 (8.1) & 6.2  (10.3)& 6.5  (11.4) & \nodata \\
\ion{Cr}{2}$\lambda$2062bl$^2$ REW (m\AA)   & 23.2 (3.8) & 31.2 (5.8) & 24.1 (6.0) & 4.6  (7.7) & 4.8  (8.5)  &\nodata  \\
\ion{Zn}{2}$\lambda$2062bl$^2$ REW (m\AA)  & 12.3 (6.2) & 26.6 (10.3)& 6.2  (9.5) & 12.4 (12.9)& 6.1  (14.3) &\nodata  \\
\ion{Cr}{2}$\lambda$2066 REW (m\AA)  & 13.0 (4.5) & 19.2 (8.3) & 9.8 (16.0) & 21.5 (10.6)& 2.5  (8.0)  &\nodata  \\
$\left[{\rm Zn}/{\rm H}\right]^{3,4}$  &$-0.88$ (0.19)&$-0.56$ (0.19)&$-0.94$ (0.21)& $<-1.08$& $<-1.05$& $-1.13$ (0.19) \\
$\left[{\rm Cr}/{\rm Zn}\right]^4$  &$-0.45$ (0.13)&$-0.64$ (0.13)&$-0.38$ (0.18)&  \nodata  & \nodata & \nodata \\
mean SDSS g mag  & 18.8 (0.6) & 18.8 (0.5) & 18.8 (0.6) & 18.8 (0.6) & 18.8 (0.7)  & \nodata \\
mean SDSS r mag  & 18.7 (0.5) & 18.7 (0.5) & 18.6 (0.6) & 18.6 (0.5) & 18.8 (0.6)  & \nodata \\
\enddata

\tablecomments{(1) The feature at $\lambda$2026 is a blend due
to \ion{Zn}{2}, \ion{Mg}{1}, and a very weak \ion{Cr}{2} line. As
measured in the composite spectra, the \ion{Mg}{1} $\lambda$2026
REW was taken to be 32.0 times smaller than the measured REW
of \ion{Mg}{2} $\lambda$2852 and the \ion{Cr}{2} $\lambda$2026
REW was taken to be 23.0 times smaller than the measured REW of
\ion{Cr}{2}$\lambda$2056. The remaining absorption was attributed to
be due to \ion{Zn}{2} $\lambda$2026. 
In a few individual
QSO spectra, comparison of the strength of \ion{Mg}{1} $\lambda$2852
to that of \ion{Mg}{2} $\lambda\lambda$2796,2803 suggests that
\ion{Mg}{1} $\lambda$2852 may be approaching saturation, which
implies that the strength of the \ion{Mg}{1} $\lambda$2852 line
may be $<32.0$ times the strength of \ion{Mg}{1} $\lambda$2026.
We mention this as a precaution which we will neglect for now.
(2) The feature at $\lambda$2062
is a blend due to \ion{Cr}{2} and \ion{Zn}{2}. The REW of \ion{Cr}{2}
$\lambda$2062 was taken to be 0.50 times the measured and summed REWs
of \ion{Cr}{2} $\lambda$2056 and \ion{Cr}{2} $\lambda$2066, or 0.75
times the measured REW of \ion{Cr}{2} $\lambda$2056, depending on
which one resulted in smaller propagated errors. The remaining absorption
was then attributed to be due to \ion{Zn}{2} $\lambda$2062. (3)
When upper limits (2$\sigma$) on metallicities are reported, they
represent those that apply for the quoted $W_0^{\lambda2796}$ limit
or interval. Since this does not include 100\% of the DLA gas (\S2.3),
a correction is made (\S3) to derive results on the cosmic DLA
gas metallicity, which is reported in the last column of this table.
(4) The following oscillator strengths are adopted: 
$f=0.4890$ for \ion{Zn}{2} $\lambda$2026.14, 
$f=0.2560$ for \ion{Zn}{2} $\lambda$2062.66, 
$f=0.00471$ for \ion{Cr}{2} $\lambda$2026.27,
$f=0.105$ for \ion{Cr}{2} $\lambda$2056.25, 
$f=0.0780$ for \ion{Cr}{2} $\lambda$2062.23, 
$f=0.0515$ for \ion{Cr}{2} $\lambda$2066.16, 
$f=0.1120$ for \ion{Mg}{1} $\lambda$2026.48, 
and $f=1.8100$ for \ion{Mg}{1} $\lambda$2852.96.
See kingpin.ucsd.edu/$\sim$hiresdla/
for details and references. Metal abundances are relative to the solar values
of Grevesse \& Sauval (1998).}
\end{deluxetable*}

\subsection{The Mean $N_{HI}$ of the Sample}

To estimate the mean neutral hydrogen column density of the absorber
sample, we have measured $N_{HI}$ (when available) for every known
\ion{Mg}{2} system with $W^{\lambda2796}_0 \ge 1$ \AA, i.e., 24
systems studied by RT2000 and 51 new systems from ongoing HST UV
spectroscopy of the Ly$\alpha$ absorption line in strong \ion{Mg}{2}
systems (Rao et al. 2003, in preparation).  These 75 systems have
neutral hydrogen column densities in the range $5\times10^{18}<$\
$N_{HI}<7\times10^{21}$ atoms cm$^{-2}$, and their mean neutral
hydrogen column density is $<$$N_{HI}$$>$ $=3.6\pm1.3\times 10^{20}$
atoms cm$^{-2}$, which we take to be the mean neutral hydrogen column 
density of our sample. There is no indication that the $<$$N_{HI}$$>$
values for the four subsamples are different from that determined
for the entire sample.

However, the REW regime considered (\S2.1) does not represent a census
of all \ion{H}{1} in the Universe over the studied redshift interval,
nor does it represent all DLAs with $N_{HI} \ge 2\times10^{20}$ atoms
cm$^{-2}$.  RT2000 estimated that 5\% of DLAs are missed using their
selection criterion and the results of Nestor et al. (2002)
indicate that about half of the DLAs are missed by excluding $0.5 <
W^{\lambda2796}_0 < 1.0$ \AA\ systems from a composite spectrum.
Our HST spectroscopy suggests that \ion{Mg}{2} systems in this
REW range have mean neutral hydrogen column density $<$$N_{HI}$$>$
$=2.7\pm1.2\times 10^{20}$ atoms cm$^{-2}$. This is 25\% smaller than
the mean neutral hydrogen column density for higher REW \ion{Mg}{2}
systems, but statistically equivalent given the measurement errors.
These effects need to be accounted for when deriving the {\it cosmic}
neutral-gas-phase metallicity of DLAs (\S3).

\section{Measurements and Abundance Results}

The basis for using the \ion{Zn}{2} and \ion{Cr}{2} absorption lines
for element abundance determinations stems from the fact that these
lines are unsaturated.  This can be verified through high-resolution
studies (e.g., Prochaska \& Wolfe 1999). Although there has been
some discussion of the importance of ionization corrections for
metal abundance determinations in DLAs (e.g., Howk \& Sembach
1999), we will not consider this since it is an issue that needs
further study and ionization corrections have not been made when
reporting DLA metal abundances in the literature. However, it is
generally agreed that the bulk of the Zn and Cr is singly ionized
in the DLA neutral regions. Thus, we assume that a determination
of [Zn$^+$/H$^0$] is equivalent to [Zn/H] and a determination of
[Cr$^+$/Zn$^+$] is equivalent to [Cr/Zn].

The notes to Table 1 provide information on the method we use
to deduce metal abundances,  but two clarifications should be
made. First, the $W_0^{\lambda2796} \ge 1.0$ \AA\ composite
over the studied redshift interval does not sample all of the
DLA gas. By excluding $W_0^{\lambda2796} < 1.0$ \AA\
systems, $\approx45$\% of the DLA gas is likely to be excluded from
the composite (\S2.3). Second, the subsample composites with $1.0
\le W_0^{\lambda2796} < 1.3$ \AA\ show no detectable metallicity
(Table 1), suggesting that the missed DLA gas has negligible
metallicity. Thus, to derive the cosmic DLA gas metallicity,
we must reduce the derived metallicity in the full composite
sample by a factor of $\approx 1.8$.  Overall we then find that
[Zn/H] = $-1.13 \pm 0.19$ in DLAs, or $7.4\pm1.1\%$ solar.
The result for Cr, which does not correct for the missed DLAs, is
[Cr/Zn] = $-0.45 \pm 0.13$, indicating that $\approx$ 65\% of the
Cr is depleted on to grains.

We also find that trends are present. The derived metallicities are
highest in the subsamples with the largest REWs, which correspond
to those with the largest velocity spreads (Table 1). Within the
high-REW regime, the Zn metallicity is larger and [Cr/Zn] is smaller
in the lower redshift composite. These results are equivalent to
column-density-weighted determinations of cosmic neutral-gas-phase
metallicities in DLAs.  Our findings are consistent with and
directly comparable to the results of Pettini et al. (1999) and,
more recently, the results summarized by Turnshek et al.  (2003).

The statistical error on $<$$N_{HI}$$>$ is 36\% (0.16 dex). Thus,
since the measured errors on REWs in the total sample and in
the two high-REW subsamples are smaller, this error dominates the
uncertainties in [Zn/H] determinations.  With the aide of future HST
UV spectroscopy of high-$N_{HI}$ systems, this error may be reduced.
The statistical errors in [Cr/Zn] determinations are smaller since
$<$$N_{HI}$$>$ does not enter into the determinations.  Recall that,
within the redshift and REW intervals under study, there is at
present no evidence that $<$$N_{HI}$$>$ is correlated with redshift
or REW. However, if such a correlation were present, it would give
rise to a systematic error requiring additional analysis to remove.
In all cases the errors quoted are statistical errors propagated
from REW and/or $<$$N_{HI}$$>$ measurement errors.

\section{Conclusions and Discussion}

We have derived cosmic DLA gas Zn and Cr abundances in two
\ion{Mg}{2} REW intervals which correspond to high and low velocity spread 
regimes. Within each of these kinematic regimes we have considered
low-redshift ($0.9 \le z < 1.35$) and high-redshift ($1.35 \le z \le 
2.0$) intervals. We find that:

(1) At redshifts $0.9 \le z \le 2.0$, after correcting for missed DLAs,
we find that the overall [Zn/H] = $-1.13 \pm 0.19$ ($\approx 7.4\%$ solar).
We find [Cr/Zn] = $-0.45 \pm 0.13$ ($\approx 65\%$ of the Cr is depleted),
but this does not correct for the missed DLAs.

(2) According to Savage \& Sembach (1996), [Cr/Zn] values in the
Milky Way Galaxy are typically $-0.5$ (halo), $-0.8$ (disk+halo),
$-1.1$ (warm disk), and $-2.1$ (cool disk).  Thus, while our measured
value of [Cr/Zn] provides evidence for significant depletion of
Cr on to dust grains, the effect is not nearly as large as in the
Galactic ISM. It is of interest to assess the affect of this dust
in the high-REW subsamples at low and high redshift since they
exhibit the highest metallicities. For the adopted $<$$N_{HI}$$>$
of our sample ($\approx 3.6 \times 10^{20}$ atoms cm$^{-2}$), this
amount of dust in the absorber rest frame would give rise to mean
values of $<$A$^{absorber}_V$$>=0.07$ mag for the low-redshift
subsample ($<$$z$$>$=1.15, [Zn/H]$=-0.56$, [Cr/Zn]$=-0.64$) and
$<$A$^{absorber}_V$$>=0.02$ mag for the high-redshift subsample
($<$$z$$>$=1.53, [Zn/H]$=-0.94$, [Cr/Zn]$=-0.38$). This is largely
independent of the nature of the dust, i.e., Galactic-like, LMC-like,
or SMC-like.  Individual high-\ion{Mg}{2}-REW absorbers would
undoubtedly exhibit a reasonable spread of A$^{absorber}_V$ values
around these mean values, and typically the low-\ion{Mg}{2}-REW
absorbers would give rise to little extinction.  For a high
REW absorber at the mean redshifts of the low and high redshift
subsamples, the mean V-band extinctions in the observer frame would
be $<$A$^{observed}_V$$>=0.16$ mag and $<$A$^{observed}_V$$>=0.06$
mag, respectively.  These results provide some indication of the
importance of selection effects caused by dust, but the possible 
exclusion of extremely dusty absorbers due to
dimming of background QSOs is still an open issue.

(3) There are clear trends in the composite spectra which indicate
that metallicities are higher in absorption systems with
larger velocity spreads (i.e., in the high-\ion{Mg}{2}-REW absorbers).
Moreover, among the high-\ion{Mg}{2}-REW absorbers, both the
metallicity and dust content increase with decreasing redshift,
with [Zn/H]$=-0.56$ (28\% solar) and [Cr/Zn]$=-0.64$ at $<$$z$$>=1.15$
and [Zn/H]$=-0.94$ (11\% solar) and  [Cr/Zn]$=-0.38$ at $<$$z$$>=1.53$.
Two explanations, or a combination of the two, seem viable. One is 
that larger velocity spreads correspond to deeper gravitational
potential wells. These more massive regions would lead to more chemically evolved
structures.  The other is that larger velocity spreads correspond to
regions with more intense bursts of star formation. This would also
lead to kinematically broad structures of chemically enriched gas.

(4) Finally, the findings presented here suggest that future analyses
of the full set of SDSS spectra will result in a significantly better
understanding of the neutral-gas-phase metallicity of the Universe
in the redshift interval $0.9 \le z \le 2.0$, due to the $\approx$
30-fold increase in the number of available SDSS spectra and the
inclusion of results on weaker \ion{Mg}{2} absorbers.

We thank members of the SDSS collaboration who made the SDSS
project a success and who made the QSO EDR spectra 
available.\footnote{Funding for the creation and distribution of 
the SDSS Archive has been provided by the Alfred P. Sloan Foundation,
the Participating Institutions, NASA, NSF, DOE, the Japanese
Monbukagakusho, and the Max Planck Society. The SDSS Web site
is http://www.sdss.org/.  The SDSS is managed by ARC for the
Participating Institutions. The Participating Institutions are
University of Chicago, Fermilab, Institute for Advanced Study,
the Japan Participation Group, Johns Hopkins University, Los
Alamos National Laboratory, Max-Planck-Institute for Astronomy,
Max-Planck-Institute for Astrophysics, New Mexico State
University, University of Pittsburgh, Princeton University,
the US Naval Observatory, and University of Washington.} We also
acknowledge support from NASA-STScI, NASA-LTSA, and NSF.  We thank
Eric Furst for initial work on detecting \ion{Mg}{2} systems in the
SDSS EDR sample of QSOs. DBN thanks Sara Ellison for discussions
on searches for dust-reddening in SDSS absorber spectra. This study
is an extension of that ongoing work.

\end{document}